\documentclass[a4paper,
               ]{jacow}
\usepackage{pdfpages,multirow,ragged2e} %
\makeatletter%
	\ifboolexpr{bool{xetex}}
	 {\renewcommand{\Gin@extensions}{.pdf,%
	                    .png,.jpg,.bmp,.pict,.tif,.psd,.mac,.sga,.tga,.gif,%
	                    .eps,.ps,%
	                    }}{}
\makeatother

\ifboolexpr{bool{xetex} or bool{luatex}} 
 {}                                      
 {\usepackage[utf8]{inputenc}}           

\usepackage[USenglish]{babel}
\usepackage{xurl}
\ifboolexpr{bool{jacowbiblatex}}%
 {%
  \addbibresource{jacow-test.bib}
  \addbibresource{biblatex-examples.bib}
 }{}
\begin{document}

\title{A life cycle assessment of the ISIS-II Neutron \\ and Muon Source}

\author{H. M. Wakeling\textsuperscript{1,}\thanks{hannah.wakeling@physics.ox.ac.uk}, John Adams Institute for Accelerator Science, University of Oxford, U.K.\\
		\textsuperscript{1}also at the ISIS Neutron and Muon Source, Rutherford Appleton Laboratory, STFC, UK}
	
\maketitle

\begin{abstract}
   The ISIS-II Neutron and Muon source is the proposed next generation of, and successor to, the ISIS Neutron and Muon Source based at the Rutherford Appleton Laboratory in the United Kingdom. Anticipated to start construction in 2031, the ISIS-II project presents a unique opportunity to incorporate environmental sustainability practices from its inception. A Life Cycle Assessment (LCA) will examine the environmental impacts associated with each of the ISIS-II design options across the stages of the ISIS-II lifecycle, encompassing construction, operation, and eventual decommissioning. This proactive approach will assess all potential areas of environmental impact and seek to identify strategies for minimizing and mitigating negative impacts, wherever feasible. This paper will provide insights into the process and first results of the LCA of the entirety of the ISIS-II project.
\end{abstract}

\section{INTRODUCTION}

Today, climate scientists report that to have a 50\% chance of keeping the Earth below the $1.5^\circ$C temperature increase from pre-industrial baselines, humanity can only emit six more years of cumulative carbon dioxide equivalent (CO$_2$e) emissions at current levels~\cite{OurWorldInData}. Global tipping points, if humanity continues as usual, could occur as soon as the 2030s\cite{GlobalTippingPoint}.

Scientific research itself has environmental impacts; many of which are largely reducible. To continue research for the overall benefit of humanity, including aiding the fight to keep the Earth habitable, the research community is swiftly altering practises and increasing efforts to reduce the direct environmental impact of research activities~\cite{HECAPplus}.

The ISIS-II Neutron and Muon Source is proposed as the UK's next generation source for the international neutron and muon user community. The ISIS-II project is currently at the feasibility and optioneering phase. This is the stage of design which has most influence on the overall environmental impact of the facility~\cite{PAS2080}.

Understanding and identifying all environmental impacts, both major and minor, is the first step in mitigating or reducing the proposed facility's environmental impact. Thus, as part of its efforts in environmental consideration, the ISIS-II project is performing a Life Cycle Assessment (LCA) to inform its design process.

\begin{figure*}[th!]
    \centering
    \includegraphics*[width=\textwidth]{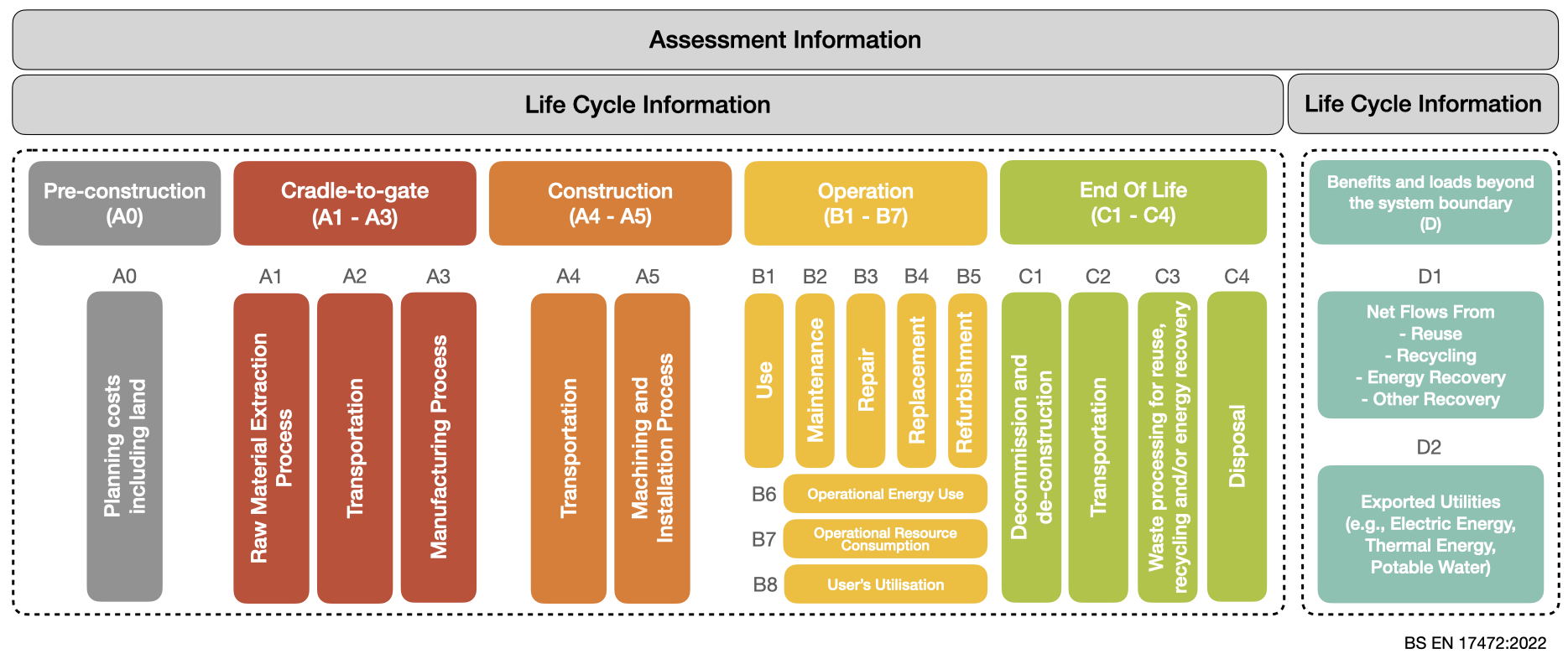}
    \caption{The BS EN17472:2022 engineering standard for building LCAs, which has been altered to be more applicable particle accelerator facilities~\cite{BSEN17472}.}
    \label{fig:BSEN17472}
\end{figure*}

\section{METHODOLOGY OF AN ISIS-II LIFE CYCLE ASSESSMENT}

An LCA is an environmental impact assessment standard. It is an iterative four-step process that examines the life cycle of an object. The four steps are:
\begin{enumerate}
    \item Goal \& Scope
    \item Inventory Analysis
    \item Life Cycle Impact Assessment (LCIA)
    \item Life Cycle Interpretation (LCI)
\end{enumerate}

To perform an LCA, resource information is required from each area of the accelerator facility, thus the methodology for performing the LCA is broken down into multiple steps. 

As the ISIS-II facility is in the feasibility and design phase, performing a simplified LCA of a still-to-be-designed facility can incur multiple sources of uncertainty at many different levels. In particular, the current estimates of materials and resources necessary for construction and operation of the ISIS-II facility have large amounts of uncertainty. Thus an iterative approach to an LCA is beneficial to the evaluation of the ISIS-II facility throughout is evolving design process.

In addition, the complexity of a full LCA performed on a large accelerator facility could be prohibitive in the progress towards results in a useful and timely manner. Thus, the LCA that will be performed will be a ``simplified" LCA. In the first instance this means that the larger impacts will be considered first, with more detail being added as it becomes available. In the second instance, the LCA will require $>$90\% of the materials and resources of the ISIS-II facility to be defined, with the remaining undefined materials and resources considered and accounted for. These undefined materials will not be ignored and due diligence will be taken to ensure that their impact is not significant with respect to the reported final results of the ISIS-II LCA.

\subsection{Defining the Goal}

The goal of the LCA is to identify the lowest lifetime environmental impact between the proposed designs of the compression ring\footnote{Each of the ISIS-II ring designs have a corresponding linear accelerator (LINAC) design necessary to inject protons at the correct energy.} for ISIS-II. 

\subsection{Deciding the Scope}

It has been proposed for the ISIS-II facility to deliver a 2.4\,MW beam of 1.2\,GeV protons to the neutron and muon community. The construction of ISIS-II is proposed to start in the 2030s. After construction, the operational lifetime of ISIS-II (2040-2100) is defined using the expected 60 year lifetime of the current ISIS neutron and muon source. The decommissioning due to radioactive isotope storage restrictions is then expected to be limited to 70 years (2100-70). Thus the functional unit of the final LCA is defined as: \textquotedblleft{One ISIS-II facility that will deliver a beam of 1.2\,GeV protons to the international neutron and muon community over a period of 60 years, with a decommissioning period of 70 years}\textquotedblright.

The ISIS-II LCA will be consider the Cradle-to-Grave life cycle of ISIS-II and its components. Fig.~\ref{fig:BSEN17472} is being used as a basis to compile the life cycle inventory.
The ReCiPe:2016\cite{ReCiPe2016} LCIA method has been chosen to evaluate the Midpoint (H) impact factors\footnote{Radionuclides from operation and the decommissioning stage may need an impact additional assessment method to be adequately evaluated.}.

To exhibit the performance of a LCA for a particle accelerator facility using open-source software and free-to-use (for academics) databases, OpenLCA software (v2.0.3~\cite{openLCA}) and the Idemat~\cite{idemat} LCA database were used for the first order-of-magnitude study. Idemat is a free database for academics and has a large selection of data available to the user evaluating a particle accelerator to the first order. In the next iteration of the LCA, this study will move to the use of the Ecoinvent database~\cite{ecoinvent}. This is due to the impact that the temporal and geographical boundaries of that free-to-use databases have on results that require more accuracy. 

Uncertainties are provided in the estimation of the inventory data\,\footnote{For a direct measurement apply: 0\%-5\% uncertainty in the data; for reliable non-measured data: 15\% uncertainty; for calculated data and extrapolations: 30\% uncertainty; for approximated data: 50\% uncertainty; and for order-of-magnitude estimates: 80\% uncertainty.} following the example from Bilan Carbone~\cite{BilanCarbone}. When performing the LCIA, a Monte Carlo simulation with a log-normal distribution evaluates the result with the uncertainty of the inventory data and data quality of the database (EcoInvent Data Quality System~\cite{ecoinvent}).

\subsection{Collecting An Inventory}

Inventory collection comprises of an evaluation of all materials and resources used in the construction, operation and decommissioning of ISIS-II. 
The design and modelling of ISIS-II is far from complete, thus the first evaluations performed for the LCA is of the components that are common in each of the designs of ISIS-II. There is less uncertainty in the designs of common components such as the Ion Source, low energy LINAC, the target and instruments. Thus this study focuses on the ISIS-II LINAC up to a proton energy of $180$\,MeV. Where designs are not yet available, existing facilities and components are used as models for the LCA:
\begin{itemize}
    \item Ion source and low energy LINAC: RAL Front End Test Stand (FETS)
    \begin{itemize}
        \item H- Ion Source
        \item Low Energy Beam Transfer (LEBT)
        \item Radio-Frequency Quadrupole (RFQ)
        \item Medium Energy Beam Transfer (MEBT)
    \end{itemize}
    \item Drift Tube LINAC (DTL): ISIS-II
    \item Separated Drift Tube LINAC (SDTL): ISIS-II
    \item LINAC shielding: ISIS-II
    \item LINAC buildings and tunnelling: ISIS-II
\end{itemize}
This includes specific additional ancillaries (klystrons, moderators, vacuum pumps), shielding, cut-and-cover tunnelling and buildings.

The operational data presented only includes UK grid electricity generation CO$_2$e emissions factor predicted up until the year 2050, which is then used as a baseline value for the remaining operational years of ISIS-II~\cite{UKGridEstimate}. Therefore, these specific operational impact results are predicted to be an overestimate, particularly due to the current efforts to de-carbonise the UK grid. The building data assumes a worst case scenario (WCS) build of 1,000\,kg\,CO$_2$e per m$^2$ of gross internal area~\cite{BuildingWCS} until further information on building suppliers and methods are available.

\section{INITIAL RESULTS}

The LCA is an iterative assessment, allowing space for improvement of data quality and, hence, uncertainties. Thus, first order-of-magnitude Global Warming Impact (GWI) from the ReCiPe:2016 LCIA results are presented for the ISIS-II facility for the low energy LINAC up to 180\,MeV. Results presented follow EN17472\cite{BSEN17472} stages A1-3 and B7. Results from the End of Life (C) study - and therefore decommissioning of ISIS-II - are not yet available.

\begin{figure}[h!]
    \centering
    \includegraphics[width=0.8\linewidth]{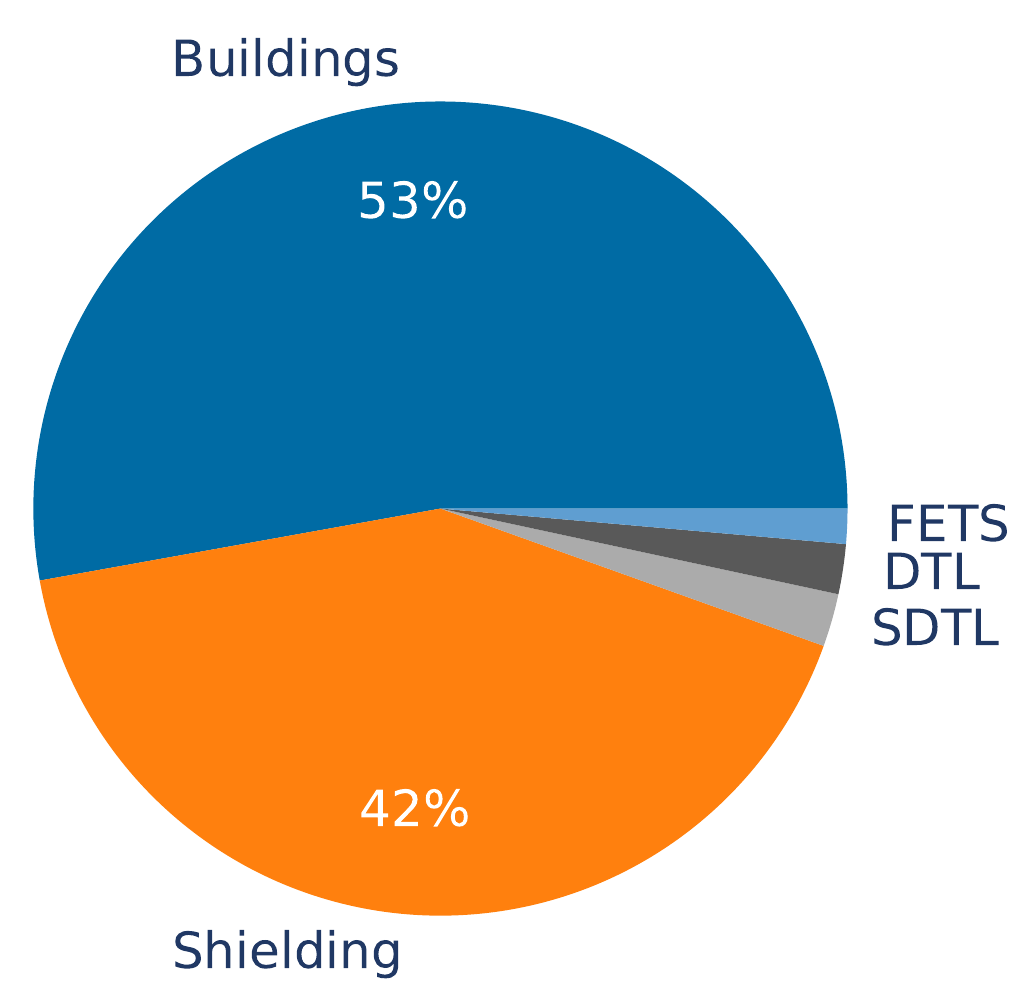}
    \caption{Distribution of the GWI for the construction materials (A1-A3) between the FETS, DTL, SDTL, shielding and buildings for the 180\,MeV LINAC.}
    \label{fig:pie_chart}
\end{figure}
It could be inferred from the order-of-magnitude result in Fig.~\ref{fig:pie_chart} that the construction materials of the ISIS-II 180\,MeV LINAC buildings and shielding structure of the accelerator could have a significantly higher environmental impactful than the materials used in the fabrication of the accelerator components in that section. In addition, the GWI of the defined ISIS-II LINAC components shows construction (A1-5) and operation (B7) to be of the same order of magnitude (Table~\ref{tab:GlobalWarmingImpact}). 
\begin{table}[!hbt]
   \centering
   \caption{GWI of the 180\,MeV LINAC for Cradle-to-Gate and Construction (A1-5) and the Lifetime Operational Energy Use (B7)}
   \begin{tabular}{lc}
       \toprule
      \textbf{Life Cycle} &  \textbf{Global Warming}  \\
      \textbf{Information} &  \textbf{[kt CO$_2$\,eq]}  \\
       \midrule
            A. Construction & $\mathcal{O}(20)$ \\
            B. Operation & $\mathcal{O}(40)$ \\
            C. Decommissioning & \small{TBC} \\
       \bottomrule
   \end{tabular}
   \label{tab:GlobalWarmingImpact}
\end{table}

This effectively provides a hotspot analysis result, indicating that our first point of concern is focusing on reducing the environmental impact of the buildings, structure and shielding of ISIS-II. 




Ultimately, this study has been able to provide a first insight into the environmental impact of a large accelerator facility, and has opened a discourse on actions to be taken into consideration when designing ISIS-II. With the expansion of the LCA to include more areas of the ISIS-II facility, inclusion of more of its life-cycle, further reduction in uncertainty of inventory estimations and access to more accurate models, the LCA will be able provide a comprehensive evaluation of the full lifetime environmental impact of the ISIS-II facility.

\section{CONCLUSION}

The first LCA results of the ISIS-II LINAC up to an energy of 180\,MeV indicates construction and operation to be of a similar order of magnitude. With the shift of the UK grid towards net-zero, it can be expected that the construction of the ISIS-II facility will have a higher GWI than its entire lifetime of operations. Investigation is underway and efforts ongoing on mitigating and reducing this environmental impact.

\section{ACKNOWLEDGEMENTS}
This work was supported by the Science and Technology Facilities Council via ST/V001655/1 and the ISIS Neutron and Muon Source. Personal thanks to Alan Letchford at RAL for providing first models of the DTL and SDTL for ISIS-II.

\ifboolexpr{bool{jacowbiblatex}}%
	{\printbibliography}%
	{%

}

\end{document}